\documentclass[aps,amssymb,amsmath,prl,reprint,noshowpacs]{revtex4-1}
\usepackage{times}
\usepackage{amssymb,amsmath,graphicx}
\usepackage[usenames]{color}
\usepackage{bbold}

\usepackage[T1]{fontenc}
\usepackage[utf8]{inputenc}
\usepackage[english]{babel}

\newcommand{\vect}[1]{\boldsymbol{#1}}
\newcommand{\abs}[1]{\left| #1 \right|} 

\newcommand{\imu}{\text{\rm i}}
\usepackage{textcomp}

\newcommand{\figwidth}{0.9\columnwidth} 
\newcommand{\commentOut}[1]{}
\usepackage{scalefnt}   

\newcommand{\affil}{Photonics Laboratory, ETH Zürich, CH-8093 Zürich, Switzerland}

\begin{document}
\scalefont{1.05}
\title{Cooling mechanical oscillators by coherent control}
\author{Martin Frimmer}
\affiliation{\affil}
\homepage{http://www.photonics.ethz.ch}
\author{Jan Gieseler}
\affiliation{\affil}
\author{Lukas Novotny}
\affiliation{\affil}

\begin{abstract}
 In optomechanics, electromagnetic fields are harnessed to control a single mode of a mechanically compliant system, while other mechanical degrees of freedom remain unaffected due to the modes' mutual orthogonality and high quality factor.
 Extension of the optical control beyond the directly addressed mode would require a controlled coupling between mechanical modes.
 Here, we introduce an optically controlled coupling between two oscillation modes of an optically levitated nanoparticle.
 We sympathetically cool one oscillation mode by coupling it coherently to the second mode, which is feedback cooled.
 Furthermore, we demonstrate coherent energy transfer between mechanical modes and discuss its application for ground-state cooling.
\end{abstract}
\date\today

\maketitle

\paragraph{Introduction.}
Coherence is at the very heart of physics and its pivotal role extends from classical to quantum physics, where the coherent evolution of a quantum mechanical wavefunction is a prerequisite for quantum coherent control~\cite{Allen1987,Mandel1995}.
With such coherent control being routinely achieved in nuclear and atomic physics~\cite{Haroche2006}, quantum engineering has emerged as a novel discipline aiming to exploit the features of quantum mechanics to outperform classical computing, sensing, and metrology~\cite{Dowling2003}.
One particularly promising testbed for quantum engineering are optomechanical systems, hybrids of a mechanical oscillator coupled to an electromagnetic field mode~\cite{Aspelmeyer2014}.
Exploiting the forces of light, single mechanical oscillator modes have been cooled from the classical realm down to their quantum ground state~\cite{Teufel2011,Chan2011}. The next step towards a quantum network is to coherently couple several such mechanical oscillators~\cite{Fang2016,Balram2016}.
One particularly interesting optomechanical system is a sub-wavelength dielectric particle levitated in a laser focus~\cite{Chang2010,Romero-Isart2011,Li2011,Gieseler2012,Vovrosh}. At sufficiently low pressures, such a particle interacts with its environment solely via the radiation field and thereby constitutes an ideal system to study both classical and quantum effects requiring minimal dephasing~\cite{Habraken2012,Schmidt2012}.
The center-of-mass motion of an optically levitated particle embodies three uncoupled harmonic oscillators and a scheme to coherently couple these different degrees of freedom would allow implementation of classical coherent control operations~\cite{Spreeuw1990,Okamoto2013, Faust2013, Okamoto2016, Xu2016}.
Such coherent control would benefit cavity-assisted cooling of a levitated nanoparticle, which holds promise to reach the quantum ground state of motion, however, merely along the cavity axis~\cite{Chang2010,Romero-Isart2011,Kiesel2013,Millen2015}. A coupling mechanism between the particle's oscillation modes would allow to transfer the cavity's cooling power to the particle's remaining degrees of freedom.
Reaching beyond applications around cooling, controlled coupling of the particle's oscillation modes would allow coherent transfer of quanta between different oscillator modes, an integral requirement for quantum coherent operations. Surprisingly, despite its potential, the controlled coupling between different degrees of freedom of an optically trapped particle has remained elusive to date.

In this Letter, we demonstrate an all-optical scheme to coherently couple two oscillation modes of a nanoparticle optically levitated in a focused laser beam.
We exploit this coupling to demonstrate two novel cooling schemes for a levitated nanoparticle. First, we sympathetically cool one oscillation mode of the particle by coupling it to a feedback-cooled mode~\cite{Larson1986,Joeckel2015}. Second, we experimentally demonstrate cooling by coherent energy transfer between oscillator modes. We establish the theoretical limit of this technique, which, in principle, allows ground state-cooling of one mode of oscillation of a levitated nanoparticle.
\begin{figure}
\includegraphics[width=\figwidth]{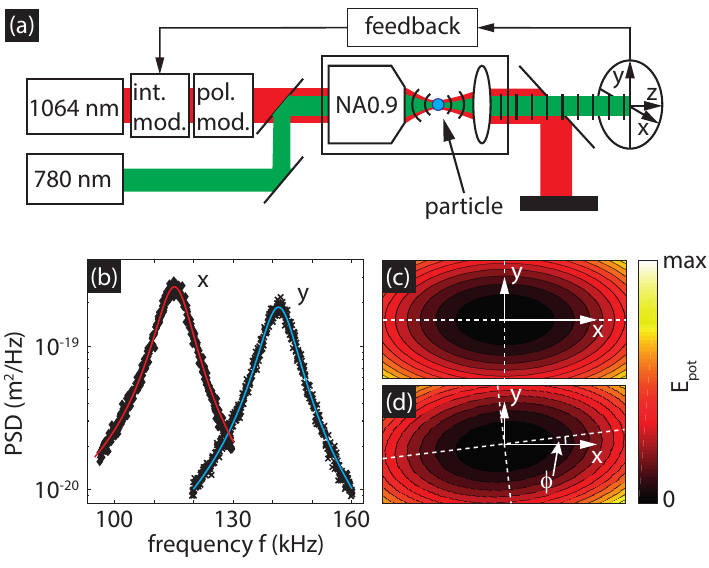}
\caption{(a)~Experimental setup. The trapping laser (1064~nm) passes two modulators for intensity and polarization modulation, before it is combined with a measurement laser (780~nm) at a dichroic beamsplitter. Both lasers are focused by a microscope objective inside a vacuum chamber and then recollimated by a lens. The trapping laser is split off by a dichroic beamsplitter and the measurement laser is guided to the detection system to measure the particle position $x,y,z$. The position measurement is used to derive a feedback signal to heat or cool the particle's motion.
(b)~Power spectral densities of particle motion along $x$ and $y$. Solid lines are Lorentzian fits.
(c)~Illustration of the optical potential in the focal plane, where the color encodes potential energy as a function of $xy$-position. The potential is stiffer along the $y$-direction due to the polarization of the trapping laser.
(d)~Illustration of the optical potential rotated by an angle $\phi$ around the optical axis.}
\label{fig:setup}
\end{figure}

\paragraph{Experimental.}
In our setup, illustrated in Fig.~\ref{fig:setup}(a), a laser beam at $1064\,\text{nm}$ ($\approx50\,\text{mW}$) is focused by an objective (NA0.9, 100x) inside a vacuum chamber. We trap silica nanospheres (diameter $136\,\text{nm}$) in the laser focus. At a dichroic beamsplitter, the trapping laser is combined with a much weaker measurement laser at $780\,\text{nm}$ ($\approx3\,\text{mW}$). A collection lens collimates both laser beams and the radiation scattered by the trapped particle. The trapping laser is filtered out at a dichroic beamsplitter while the measurement beam is guided to the detection optics, where the particle position is measured, as described in Ref.~\onlinecite{Gieseler2012}. In our notation, $z$ is the direction along the optical axis and  $x(y)$ points along the horizontal (vertical) axis in the focal plane. Throughout this paper, we focus on the control of the particle's motion in the $xy$-plane.
To first order, the optical potential is harmonic around the laser focus and our trapped particle therefore resembles three uncoupled harmonic oscillators. Accordingly, for each axis, the power spectral density (PSD) of the particle trajectory is a Lorentzian. Figure~\ref{fig:setup}(b) shows the power spectra of the motion in the focal plane of a particle trapped at a pressure of $10\,\text{mbar}$. The eigenfrequency of the $x$-mode ($y$-mode) is $\Omega_x = 2\pi\cdot115\,\text{kHz}$ ($\Omega_y = 2\pi\cdot141\,\text{kHz}$) and can be tuned via the power of the trapping laser. The frequency difference between the $x$- and $y$-modes arises from the asymmetry of the trapping potential due to the trapping laser being linearly polarized along the $x$-axis, as illustrated in Fig.~\ref{fig:setup}(c). The width of each Lorentzian in the PSD is set by the damping rate $\gamma$. While at extremely low pressures the damping is governed by radiation pressure shot noise of the trapping laser~\cite{Jain2016}, throughout this paper, unless noted otherwise, we work at $5\cdot10^{-6}\,\text{mbar}$ where coupling to the gas in the chamber dominates the damping. Thus, according to the equipartition theorem, the area under the PSD of each oscillator in thermal equilibrium has to equal $k_BT_0/(m\Omega^2)$, where $k_B$ is Boltzmann's constant, $T_0$ denotes room temperature, and $m$ is the particle mass, allowing us to convert our detector signals from volts to meters.
Our setup is equipped with a feedback mechanism, enabling us to individually control the energy in each oscillation mode of the particle by parametrically modulating the trap stiffness via the trapping-laser intensity~\cite{Gieseler2012}. By adjusting the phase of the feedback signal, the oscillation amplitude of the particle can both be reduced (parametric cooling) or increased (parametric heating). Note that throughout this paper, the terms \emph{heating} and \emph{cooling} refer to the particle's center-of-mass motion.

We couple the $x$- and $y$-motion of the trapped particle by harmonically modulating the polarization angle $\phi$ of the trapping laser with an electro-optic modulator (EOM) according to $\phi(t)=\phi_0\cos(\omega_\text{mod} t)$, where $\phi_0$ is proportional to the voltage $V_0$ applied to the EOM. This modulation corresponds to a rotation of the optical potential, as illustrated in Fig.~\ref{fig:setup}(d). Counterintuitively, while introduction of a static rotation angle $\phi$ leaves the eigenmodes of the oscillator system unchanged, harmonically modulating the polarization angle $\phi$ (with $\phi\ll1$) generates a coupling between the eigenmodes.

\paragraph{Coherent control of two-mode system.}
\begin{figure}
\includegraphics[width=\columnwidth]{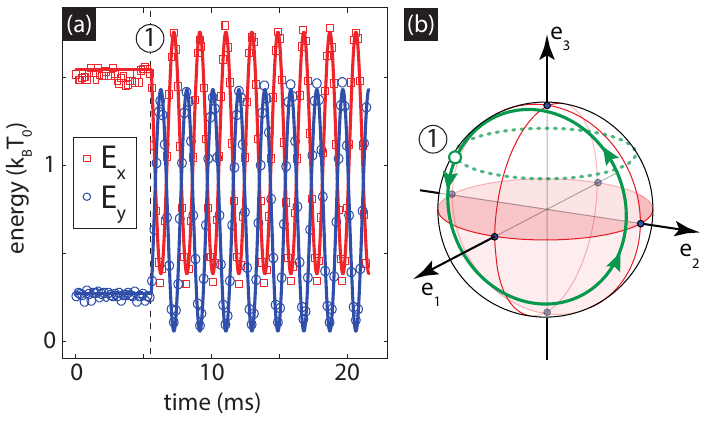}
\caption{
(a)~Rabi oscillations of mode energy. The particle is initialized in a state with high energy in the $x$-mode by parametric heating, while the $y$-mode is feedback cooled. At time $t=0$ the feedback is turned off and remains off during the entire measurement. At time $t=5.5$~ms [denoted as (1)], we turn on the modulation of the polarization angle of the trapping beam and observe periodic exchange of energy between the modes. Symbols denote measurements, solid lines theory.
(b)~Bloch sphere representation of measurement in (a). At $t=0$ the system is initialized to a point on the green dashed line on the upper hemisphere, with the precise location determined by the phase between the oscillator modes, which is not explicitly controlled experimentally. When the modulation is switched on, the system in the measurement in (a) is at the point denoted as (1), and the Bloch vector of the system rotates along the solid green line.
}
\label{fig:Rabi_principle}
\end{figure}
%
We demonstrate our mode-coupling scheme in Fig.~\ref{fig:Rabi_principle}(a), where we plot the energies $E_x$ and $E_y$ carried by the $x$- and $y$-mode, respectively, as a function of time. Using parametric heating/cooling, we initialize the $x$-mode ($y$-mode) of the particle with an energy $E_x=1.5\,k_BT_0$ ($E_y=0.25\,k_BT_0$). We stress that during the entire time shown in Fig.~\ref{fig:Rabi_principle}(a) any feedback is switched off. After $5.5\,\text{ms}$ [point (1) in Fig.~\ref{fig:Rabi_principle}(a)], we switch on the polarization modulation with an amplitude of $V_0=300\,\text{mV}$ and a frequency $\omega_\text{mod}$ close to the frequency difference of the eigenmodes $\Delta\Omega=\Omega_y-\Omega_x=26\,\text{kHz}$. We observe a periodic exchange of energy between the two modes.

To understand our data, we describe our system in a slowly varying envelope approximation, where $\bar{a}(t)$ and $\bar{b}(t)$ are the complex amplitudes of the $x$- and $y$-modes, respectively, in a frame rotating at the frequency of the modulation voltage $\omega_\text{mod}$.
Neglecting thermal forces, the classical equations of motion of these mode amplitudes read~\cite{Frimmer2014}
\begin{equation}
\label{eq:SVEAsystemRotFrame}
\imu\begin{bmatrix} \dot{\bar{a}} \\ \dot{\bar{b}}\end{bmatrix} \;=\; \frac{1}{2}\begin{bmatrix} \delta-\imu\gamma \,& \;-A\\  -A \,& -\delta-\imu\gamma \end{bmatrix} \begin{bmatrix} \bar{a}\\\bar{b}\end{bmatrix},
\end{equation}
where $\delta=\omega_\text{mod}-\Delta\Omega$ denotes the detuning of the modulation frequency $\omega_\text{mod}$ from the bare-mode frequency difference $\Delta\Omega$ and the coupling rate $A=\phi_0\Delta\Omega$ is proportional to the modulation angle of the optical potential and the bare-mode frequency difference.
Multiplying both sides of Eq.~\eqref{eq:SVEAsystemRotFrame} with $\hbar$ brings the classical equations of motion into the shape of the Schrödinger equation of a quantum mechanical two-level system. The complex amplitudes $\bar{a}(t)$ and $\bar{b}(t)$ represent the complex amplitudes of the excited and ground state, respectively, and the level populations $\abs{\bar{a}(t)}^2$ and $\abs{\bar{b}(t)}^2$ are proportional to the energies $E_x$ and $E_y$ in the respective modes of oscillation.

According to Eq.~\eqref{eq:SVEAsystemRotFrame}, the frequency of the energy exchange between the $x$- and $y$-modes, as observed in Fig.~\ref{fig:Rabi_principle}(a), is given by the generalized Rabi frequency $\Omega_R=\sqrt{A^2+\delta^2}$. The solid lines in Fig.~\ref{fig:Rabi_principle}(a) show the solutions of Eq.~\eqref{eq:SVEAsystemRotFrame} and are in good agreement with the measurement.
Like any two-mode system, quantum or classical, our system can be described on the Bloch sphere, sketched in Fig.~\ref{fig:Rabi_principle}(b)~\cite{Okamoto2013,Faust2013,Frimmer2014}. When all energy resides in the $x$-mode ($y$-mode), the system is located at the north pole (south pole) and the amplitude $\bar{b}$ ($\bar{a}$) vanishes. Points on the equator denote states with equal energy in both modes. In the measurement in Fig.~\ref{fig:Rabi_principle}(a), our system is initialized in a state on the upper Bloch hemisphere on a circle of constant latitude [green dashed line in Fig.~\ref{fig:Rabi_principle}(b)], the precise position along this circle being determined by the relative phase between the two modes, which we do not explicitly control experimentally. When the coupling between the modes is switched on, and the modulation frequency is on resonance, the Bloch vector of the system rotates around the $\vect{e}_1$-axis in Bloch space. In the measurement in Fig.~\ref{fig:Rabi_principle}(a), when the coupling is turned on, the system is at the point denoted with (1) in Fig.~\ref{fig:Rabi_principle}(b), such that the Bloch vector rotates on the trajectory denoted as the green solid line in Fig.~\ref{fig:Rabi_principle}(b).

\paragraph{Sympathetic cooling.}
\begin{figure}
\includegraphics[width=\figwidth]{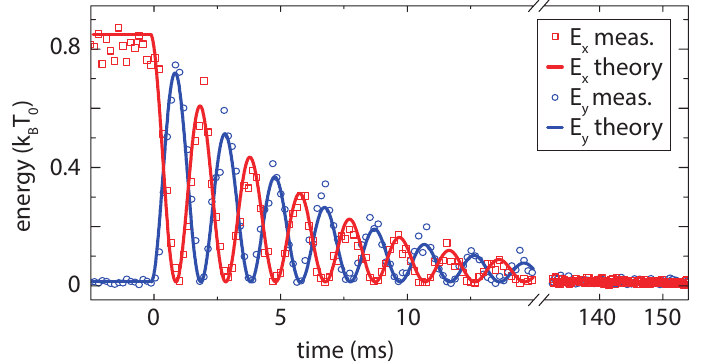}
\caption{Sympathetic cooling of the $x$-mode. The system is started with the coupling off and the $y$-mode cooled to $0.01\,kT_0$. The $y$-mode is under feedback cooling during the entire time shown. No feedback cooling is applied to the $x$-mode throughout the entire measurement. The initial energy of the $x$-mode is $0.8\,kT_0$. At time $t=0$, the coupling between the $x$- and $y$-modes is turned on and energy is transferred from the hot $x$-mode to the $y$-mode, from where it is removed by feedback cooling, leading to sympathetic cooling of the $x$-mode.}
\label{fig:sympCooling}
\end{figure}
In analogy to atomic physics, where sympathetic cooling is a well established technique to cool degrees of freedom inaccessible to direct laser cooling~\cite{Larson1986}, we introduce a coherent control scheme to sympathetically cool one oscillation mode of the levitated nanoparticle by coupling it to the other oscillation mode, which is feedback-cooled. Figure~\ref{fig:sympCooling} shows our experimental results.
The symbols denote the measured energy in the $x$- and $y$-modes of the levitated particle as a function of time and the solid lines are analytical solutions according to Eq.~\eqref{eq:SVEAsystemRotFrame}. Throughout the entire measurement, we feedback cool the $y$-motion of the particle while its $x$-motion is freely evolving. At the beginning of the measurement, the $x$-mode carries $0.8\,k_BT_0$ of energy, while the $y$-mode is feedback-cooled to $0.01\,k_BT_0$. At time $t=0$, we switch on the coupling between the two modes and observe the characteristic Rabi oscillations. Strikingly, however, the energy in both modes decays exponentially. Clearly, the coupling transfers energy from the uncooled $x$-mode to the $y$-mode, from which the energy is removed by feedback cooling. The decay time of the envelope of the mode populations in Fig.~\ref{fig:sympCooling} is set by the cooling rate of the feedback applied to the $y$-mode. For times significantly longer than the inverse cooling rate, both modes approach the steady-state temperature of the feedback-cooled mode. We point out that while we have measured the position of the $x$-oscillator here in order to demonstrate sympathetic cooling, the only information required about the $x$-mode in this scheme is its eigenfrequency, in order to adjust the modulation frequency correctly.
Thus, it is not necessary to monitor the temporal evolution of the two modes.
The $x$-mode eigenfrequency can be extracted by exclusively monitoring the $y$-mode's population while sweeping the coupling frequency, and, consequently, an entirely ``dark'' mode, inaccessible by optical detection, can be cooled by our sympathetic cooling scheme.

\paragraph{Energy-transfer cooling.}
\begin{figure}
\includegraphics[width=\figwidth]{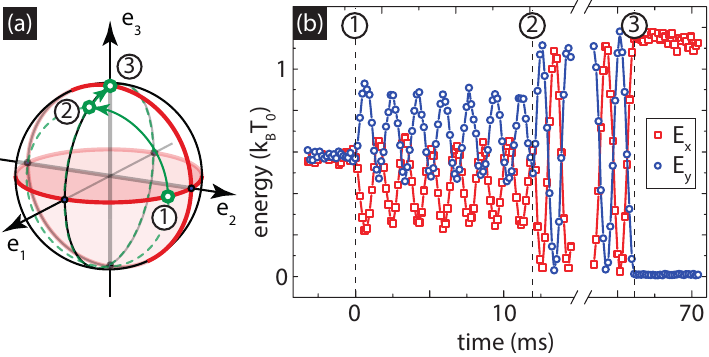}
\caption{(a)~Protocol for cooling by energy transfer. The system is started at an arbitrary point on the Bloch sphere, here chosen to be the point on the equator denoted by (1). Coupling between the modes rotates the Bloch vector around the $\vect{e}_1$-axis. Measuring the phase of the Rabi oscillations allows prediction of the passage through point (2), where the phase of the modulation signal is switched, such that the Bloch vector rotates around the $\vect{e}_2$-axis. Again, the Rabi oscillations are monitored to extract their phase and predict passage through the pole of the Bloch sphere, denoted by (3), where the modulation is switched off.
(b)~Experimental implementation of energy-transfer protocol. The system is brought from point (1), with equal energy of $0.6\,k_BT_0$ in both modes, to a state where the $y$-mode carries $0.01\,k_BT_0$. Datapoints are measured, solid lines serve as guides to the eye.}
\label{fig:energyTransfer}
\end{figure}
The final temperature reachable by sympathetic cooling is determined by the temperature of the cold bath, which, in our case, corresponds to the temperature of the mode that is actively cooled by feedback. We now introduce a coherent energy-transfer protocol, which, in principle, allows one of the modes to be cooled to the quantum ground-state.
The scheme is illustrated on the Bloch sphere in Fig.~\ref{fig:energyTransfer}(a). The objective is to transfer the system from an arbitrary starting point to the north pole of the Bloch sphere, where the energy of the $y$-mode is at the zero-point level  and all remaining energy is contained in the $x$-mode. Without loss of generality, we assume as the starting point an arbitrary point on the equator, denoted by (1) in Fig.~\ref{fig:energyTransfer}(a), corresponding to equal energy in both modes and arbitrary phase between them.
Turning on the coupling between the modes rotates the Bloch vector around the $\vect{e}_1$-axis along the dashed green line. The mode energies are constantly monitored for several Rabi cycles to determine the phase of the Rabi oscillations, such that the passage of the Bloch vector through the $\vect{e}_1\vect{e}_3$-plane in Bloch space can be predicted [point (2) in Fig.~\ref{fig:energyTransfer}(a)]. When this point is reached, the phase of the modulation signal is switched by $\pi/2$, such that the Bloch vector now rotates around the $\vect{e}_2$-axis. Again, the Rabi oscillations are monitored and their phase is determined, in order to predict the point of passage through the pole of the Bloch sphere [point (3) in Fig.~\ref{fig:energyTransfer}(a)], where the coupling is switched off.

Figure~\ref{fig:energyTransfer}(b) shows an experimental implementation of our protocol. We initialize our system with equal energy of $0.6~k_BT_0$ in both the $x$- and the $y$-mode. Note that the feedback is switched off throughout the entire experiment. At time $t=0$, we turn on the coupling between the modes, observe several Rabi cycles and determine their phase \emph{in situ}, in order to switch the phase of the modulation signal by $\pi/2$ at about $12\,\text{ms}$, corresponding to point (2) of our protocol.
We note that due to a slight detuning of the modulation field, the rotation vector has a small $\vect{e}_3$-component between points (1) and (2), giving rise to slightly asymmetric oscillations of the energy in the $x$- and $y$-modes. This fact does not affect our cooling scheme.
After switching the phase of the modulation signal, we again monitor the Rabi oscillations, in order to switch off the coupling between the modes at point (3). At the end of the experiment, the $y$-mode is cooled to an energy of $0.01\,k_BT_0$, while the $x$-mode carries all remaining energy, which is roughly $1.2\,k_BT_0$.

The limit of cooling by energy transfer is set by the precision with which the phase of the Rabi oscillations of the mode energy can be determined, which in turn depends on the precision of the measurement of the particle's position, given by the noise power spectral density $S_x^\text{noise}$. Any quadrature of a time harmonic signal $u(t)$ sampled $N$ times with a variance $\sigma_u^2$ can be determined with a variance $\sigma^2_\text{quad}=2\sigma_u^2/N$. We therefore find that cooling by energy transfer is limited to a minimal mode energy of $\langle E_\mathrm{min}\rangle=\frac{1}{2}m\Omega^2 S_{x}^\text{noise}/\tau$, with $m$ the oscillator mass, $\Omega$ its eigenfrequency, and $\tau$ the time over which the Rabi oscillations are observed. In the absence of pure dephasing due to frequency fluctuations of the oscillator modes, the upper limit of $\tau$ is set by the inverse intrinsic linewidth $\gamma$. For shot noise limited detection of particle position in ultra-high vacuum, the noise floor amounts to $S_{x}^\text{noise}=1\,\text{pm}^2/\text{Hz}$ and the quality factor of the particle's oscillation modes assumes values of $Q\sim10^{9}$~\cite{Jain2016}. Based on these values, we calculate a minimum center-of-mass temperature in the pico-Kelvin range, which is significantly below the ground-state temperature ($\sim6\,\mu \text{K}$) of our nanoparticle.  Hence, we conclude that the ground state can be reached for one oscillation mode of an optically levitated nanoparticle using energy-transfer cooling.

We point out that our cooling scheme based on energy transfer reaches beyond levitated nanoparticles. Virtually all mechanical systems used in the context of optomechanics feature a plethora of modes, which are typically all ignored except one, which is addressed optically. With a coherent coupling scheme interfacing these modes, it is possible to selectively cool certain modes by transferring their energy to other modes serving as ``energy buffers'', and for which a weak coupling to a cold reservoir is sufficient.

In conclusion, we have demonstrated all-optical control of the coupling between different oscillation modes of an optically levitated nanoparticle. Our coupling scheme provides a means to coherently transfer energy between the coupled modes of oscillation. With levitated nanoparticles approaching populations of single quanta in their center-of-mass oscillation modes, our scheme will allow coherent transfer of single excitations between different oscillator modes of the same particle. Furthermore, cavity-based cooling schemes for levitated nanoparticles are expected to outperfom feedback cooling in the near future for particle motion along the cavity axis~\cite{Kiesel2013,Millen2015}. Our sympathetic cooling scheme opens up the possibility to transfer the cavity's cooling power to the particle's remaining degrees of freedom and eventually reach the quantum ground-state of all oscillation modes.

\begin{acknowledgments}
This research was supported by ERC-QMES (Grant No. 338763) and the NCCR-QSIT program (Grant No. 51NF40-160591). J.G.\ has been partially supported by the Postdoc-Program of the German Academic Exchange Service (DAAD) and H2020-MSCA-IF-2014 under REA grant Agreement No.\ 655369. We thank L.\ Rondin, E.\ Hebestreit, V.\ Jain and R.\ Reimann for valuable input and discussions.
\end{acknowledgments}

\bibliography{FrimmerRabiCoolingParticle}

\begin{thebibliography}{27}%
\makeatletter
\providecommand \@ifxundefined [1]{%
 \@ifx{#1\undefined}
}%
\providecommand \@ifnum [1]{%
 \ifnum #1\expandafter \@firstoftwo
 \else \expandafter \@secondoftwo
 \fi
}%
\providecommand \@ifx [1]{%
 \ifx #1\expandafter \@firstoftwo
 \else \expandafter \@secondoftwo
 \fi
}%
\providecommand \natexlab [1]{#1}%
\providecommand \enquote  [1]{``#1''}%
\providecommand \bibnamefont  [1]{#1}%
\providecommand \bibfnamefont [1]{#1}%
\providecommand \citenamefont [1]{#1}%
\providecommand \href@noop [0]{\@secondoftwo}%
\providecommand \href [0]{\begingroup \@sanitize@url \@href}%
\providecommand \@href[1]{\@@startlink{#1}\@@href}%
\providecommand \@@href[1]{\endgroup#1\@@endlink}%
\providecommand \@sanitize@url [0]{\catcode `\\12\catcode `\$12\catcode
  `\&12\catcode `\#12\catcode `\^12\catcode `\_12\catcode `\%12\relax}%
\providecommand \@@startlink[1]{}%
\providecommand \@@endlink[0]{}%
\providecommand \url  [0]{\begingroup\@sanitize@url \@url }%
\providecommand \@url [1]{\endgroup\@href {#1}{\urlprefix }}%
\providecommand \urlprefix  [0]{URL }%
\providecommand \Eprint [0]{\href }%
\providecommand \doibase [0]{http://dx.doi.org/}%
\providecommand \selectlanguage [0]{\@gobble}%
\providecommand \bibinfo  [0]{\@secondoftwo}%
\providecommand \bibfield  [0]{\@secondoftwo}%
\providecommand \translation [1]{[#1]}%
\providecommand \BibitemOpen [0]{}%
\providecommand \bibitemStop [0]{}%
\providecommand \bibitemNoStop [0]{.\EOS\space}%
\providecommand \EOS [0]{\spacefactor3000\relax}%
\providecommand \BibitemShut  [1]{\csname bibitem#1\endcsname}%
\let\auto@bib@innerbib\@empty
\bibitem [{\citenamefont {Allen}\ and\ \citenamefont
  {Eberly}(1987)}]{Allen1987}%
  \BibitemOpen
  \bibfield  {author} {\bibinfo {author} {\bibfnamefont {L.}~\bibnamefont
  {Allen}}\ and\ \bibinfo {author} {\bibfnamefont {J.~H.}\ \bibnamefont
  {Eberly}},\ }\href {http://books.google.ch/books?id=1q0ae-XNmWwC} {\emph
  {\bibinfo {title} {Optical Resonance and Two-level Atoms}}},\ Dover Books on
  Physics Series\ (\bibinfo  {publisher} {Dover},\ \bibinfo {year}
  {1987})\BibitemShut {NoStop}%
\bibitem [{\citenamefont {Mandel}\ and\ \citenamefont
  {Wolf}(1995)}]{Mandel1995}%
  \BibitemOpen
  \bibfield  {author} {\bibinfo {author} {\bibfnamefont {L.}~\bibnamefont
  {Mandel}}\ and\ \bibinfo {author} {\bibfnamefont {E.}~\bibnamefont {Wolf}},\
  }\href {https://books.google.ch/books?id=FeBix14iM70C} {\emph {\bibinfo
  {title} {Optical Coherence and Quantum Optics}}}\ (\bibinfo  {publisher}
  {Cambridge University Press},\ \bibinfo {year} {1995})\BibitemShut {NoStop}%
\bibitem [{\citenamefont {Haroche}\ and\ \citenamefont
  {Raimond}(2006)}]{Haroche2006}%
  \BibitemOpen
  \bibfield  {author} {\bibinfo {author} {\bibfnamefont {S.}~\bibnamefont
  {Haroche}}\ and\ \bibinfo {author} {\bibfnamefont {J.}~\bibnamefont
  {Raimond}},\ }\href {https://books.google.ch/books?id=QY6YuU-Qi-AC} {\emph
  {\bibinfo {title} {Exploring the Quantum: Atoms, Cavities, and Photons}}},\
  Oxford Graduate Texts\ (\bibinfo  {publisher} {OUP Oxford},\ \bibinfo {year}
  {2006})\BibitemShut {NoStop}%
\bibitem [{\citenamefont {Dowling}\ and\ \citenamefont
  {Milburn}(2003)}]{Dowling2003}%
  \BibitemOpen
  \bibfield  {author} {\bibinfo {author} {\bibfnamefont {J.~P.}\ \bibnamefont
  {Dowling}}\ and\ \bibinfo {author} {\bibfnamefont {G.~J.}\ \bibnamefont
  {Milburn}},\ }\href {\doibase 10.1098/rsta.2003.1227} {\bibfield  {journal}
  {\bibinfo  {journal} {Phil. Trans. R. Soc. A}\ }\textbf {\bibinfo {volume}
  {361}},\ \bibinfo {pages} {1655} (\bibinfo {year} {2003})}\BibitemShut
  {NoStop}%
\bibitem [{\citenamefont {Aspelmeyer}\ \emph {et~al.}(2014)\citenamefont
  {Aspelmeyer}, \citenamefont {Kippenberg},\ and\ \citenamefont
  {Marquardt}}]{Aspelmeyer2014}%
  \BibitemOpen
  \bibfield  {author} {\bibinfo {author} {\bibfnamefont {M.}~\bibnamefont
  {Aspelmeyer}}, \bibinfo {author} {\bibfnamefont {T.~J.}\ \bibnamefont
  {Kippenberg}}, \ and\ \bibinfo {author} {\bibfnamefont {F.}~\bibnamefont
  {Marquardt}},\ }\href {\doibase 10.1103/RevModPhys.86.1391} {\bibfield
  {journal} {\bibinfo  {journal} {Rev. Mod. Phys.}\ }\textbf {\bibinfo {volume}
  {86}},\ \bibinfo {pages} {1391} (\bibinfo {year} {2014})}\BibitemShut
  {NoStop}%
\bibitem [{\citenamefont {Teufel}\ \emph {et~al.}(2011)\citenamefont {Teufel},
  \citenamefont {Donner}, \citenamefont {Li}, \citenamefont {Harlow},
  \citenamefont {Allman}, \citenamefont {Cicak}, \citenamefont {Sirois},
  \citenamefont {Whittaker}, \citenamefont {Lehnert},\ and\ \citenamefont
  {Simmonds}}]{Teufel2011}%
  \BibitemOpen
  \bibfield  {author} {\bibinfo {author} {\bibfnamefont {J.~D.}\ \bibnamefont
  {Teufel}}, \bibinfo {author} {\bibfnamefont {T.}~\bibnamefont {Donner}},
  \bibinfo {author} {\bibfnamefont {D.}~\bibnamefont {Li}}, \bibinfo {author}
  {\bibfnamefont {J.~W.}\ \bibnamefont {Harlow}}, \bibinfo {author}
  {\bibfnamefont {M.~S.}\ \bibnamefont {Allman}}, \bibinfo {author}
  {\bibfnamefont {K.}~\bibnamefont {Cicak}}, \bibinfo {author} {\bibfnamefont
  {A.~J.}\ \bibnamefont {Sirois}}, \bibinfo {author} {\bibfnamefont {J.~D.}\
  \bibnamefont {Whittaker}}, \bibinfo {author} {\bibfnamefont {K.~W.}\
  \bibnamefont {Lehnert}}, \ and\ \bibinfo {author} {\bibfnamefont {R.~W.}\
  \bibnamefont {Simmonds}},\ }\href {\doibase 10.1038/nature10261} {\bibfield
  {journal} {\bibinfo  {journal} {Nature}\ }\textbf {\bibinfo {volume} {475}},\
  \bibinfo {pages} {359} (\bibinfo {year} {2011})}\BibitemShut {NoStop}%
\bibitem [{\citenamefont {Chan}\ \emph {et~al.}(2011)\citenamefont {Chan},
  \citenamefont {Alegre}, \citenamefont {Safavi-Naeini}, \citenamefont {Hill},
  \citenamefont {Krause}, \citenamefont {Gr{\"o}blacher}, \citenamefont
  {Aspelmeyer},\ and\ \citenamefont {Painter}}]{Chan2011}%
  \BibitemOpen
  \bibfield  {author} {\bibinfo {author} {\bibfnamefont {J.}~\bibnamefont
  {Chan}}, \bibinfo {author} {\bibfnamefont {T.~M.}\ \bibnamefont {Alegre}},
  \bibinfo {author} {\bibfnamefont {A.~H.}\ \bibnamefont {Safavi-Naeini}},
  \bibinfo {author} {\bibfnamefont {J.~T.}\ \bibnamefont {Hill}}, \bibinfo
  {author} {\bibfnamefont {A.}~\bibnamefont {Krause}}, \bibinfo {author}
  {\bibfnamefont {S.}~\bibnamefont {Gr{\"o}blacher}}, \bibinfo {author}
  {\bibfnamefont {M.}~\bibnamefont {Aspelmeyer}}, \ and\ \bibinfo {author}
  {\bibfnamefont {O.}~\bibnamefont {Painter}},\ }\href@noop {} {\bibfield
  {journal} {\bibinfo  {journal} {Nature}\ }\textbf {\bibinfo {volume} {478}},\
  \bibinfo {pages} {89} (\bibinfo {year} {2011})}\BibitemShut {NoStop}%
\bibitem [{\citenamefont {Fang}\ \emph {et~al.}(2016)\citenamefont {Fang},
  \citenamefont {Matheny}, \citenamefont {Luan},\ and\ \citenamefont
  {Painter}}]{Fang2016}%
  \BibitemOpen
  \bibfield  {author} {\bibinfo {author} {\bibfnamefont {K.}~\bibnamefont
  {Fang}}, \bibinfo {author} {\bibfnamefont {M.~H.}\ \bibnamefont {Matheny}},
  \bibinfo {author} {\bibfnamefont {X.}~\bibnamefont {Luan}}, \ and\ \bibinfo
  {author} {\bibfnamefont {O.}~\bibnamefont {Painter}},\ }\href@noop {}
  {\bibfield  {journal} {\bibinfo  {journal} {Nat. Photon.}\ }\textbf {\bibinfo
  {volume} {10}},\ \bibinfo {pages} {489} (\bibinfo {year} {2016})}\BibitemShut
  {NoStop}%
\bibitem [{\citenamefont {Balram}\ \emph {et~al.}(2016)\citenamefont {Balram},
  \citenamefont {Davan\c{c}o}, \citenamefont {Song},\ and\ \citenamefont
  {Srinivasan}}]{Balram2016}%
  \BibitemOpen
  \bibfield  {author} {\bibinfo {author} {\bibfnamefont {K.~C.}\ \bibnamefont
  {Balram}}, \bibinfo {author} {\bibfnamefont {M.~I.}\ \bibnamefont
  {Davan\c{c}o}}, \bibinfo {author} {\bibfnamefont {J.~D.}\ \bibnamefont
  {Song}}, \ and\ \bibinfo {author} {\bibfnamefont {K.}~\bibnamefont
  {Srinivasan}},\ }\href@noop {} {\bibfield  {journal} {\bibinfo  {journal}
  {Nat. Photon.}\ }\textbf {\bibinfo {volume} {10}},\ \bibinfo {pages} {346}
  (\bibinfo {year} {2016})}\BibitemShut {NoStop}%
\bibitem [{\citenamefont {Chang}\ \emph {et~al.}(2010)\citenamefont {Chang},
  \citenamefont {Regal}, \citenamefont {Papp}, \citenamefont {Wilson},
  \citenamefont {Ye}, \citenamefont {Painter}, \citenamefont {Kimble},\ and\
  \citenamefont {Zoller}}]{Chang2010}%
  \BibitemOpen
  \bibfield  {author} {\bibinfo {author} {\bibfnamefont {D.~E.}\ \bibnamefont
  {Chang}}, \bibinfo {author} {\bibfnamefont {C.~A.}\ \bibnamefont {Regal}},
  \bibinfo {author} {\bibfnamefont {S.~B.}\ \bibnamefont {Papp}}, \bibinfo
  {author} {\bibfnamefont {D.~J.}\ \bibnamefont {Wilson}}, \bibinfo {author}
  {\bibfnamefont {J.}~\bibnamefont {Ye}}, \bibinfo {author} {\bibfnamefont
  {O.}~\bibnamefont {Painter}}, \bibinfo {author} {\bibfnamefont {H.~J.}\
  \bibnamefont {Kimble}}, \ and\ \bibinfo {author} {\bibfnamefont
  {P.}~\bibnamefont {Zoller}},\ }\href {\doibase 10.1073/pnas.0912969107}
  {\bibfield  {journal} {\bibinfo  {journal} {Proc. Natl. Acad. Sci. USA}\
  }\textbf {\bibinfo {volume} {107}},\ \bibinfo {pages} {1005} (\bibinfo {year}
  {2010})}\BibitemShut {NoStop}%
\bibitem [{\citenamefont {Romero-Isart}\ \emph {et~al.}(2011)\citenamefont
  {Romero-Isart}, \citenamefont {Pflanzer}, \citenamefont {Juan}, \citenamefont
  {Quidant}, \citenamefont {Kiesel}, \citenamefont {Aspelmeyer},\ and\
  \citenamefont {Cirac}}]{Romero-Isart2011}%
  \BibitemOpen
  \bibfield  {author} {\bibinfo {author} {\bibfnamefont {O.}~\bibnamefont
  {Romero-Isart}}, \bibinfo {author} {\bibfnamefont {A.~C.}\ \bibnamefont
  {Pflanzer}}, \bibinfo {author} {\bibfnamefont {M.~L.}\ \bibnamefont {Juan}},
  \bibinfo {author} {\bibfnamefont {R.}~\bibnamefont {Quidant}}, \bibinfo
  {author} {\bibfnamefont {N.}~\bibnamefont {Kiesel}}, \bibinfo {author}
  {\bibfnamefont {M.}~\bibnamefont {Aspelmeyer}}, \ and\ \bibinfo {author}
  {\bibfnamefont {J.~I.}\ \bibnamefont {Cirac}},\ }\href {\doibase
  10.1103/PhysRevA.83.013803} {\bibfield  {journal} {\bibinfo  {journal} {Phys.
  Rev. A}\ }\textbf {\bibinfo {volume} {83}},\ \bibinfo {pages} {013803}
  (\bibinfo {year} {2011})}\BibitemShut {NoStop}%
\bibitem [{\citenamefont {Li}\ \emph {et~al.}(2011)\citenamefont {Li},
  \citenamefont {Kheifets},\ and\ \citenamefont {Raizen}}]{Li2011}%
  \BibitemOpen
  \bibfield  {author} {\bibinfo {author} {\bibfnamefont {T.}~\bibnamefont
  {Li}}, \bibinfo {author} {\bibfnamefont {S.}~\bibnamefont {Kheifets}}, \ and\
  \bibinfo {author} {\bibfnamefont {M.~G.}\ \bibnamefont {Raizen}},\ }\href
  {\doibase 10.1038/nphys1952} {\bibfield  {journal} {\bibinfo  {journal}
  {Nature Phys.}\ }\textbf {\bibinfo {volume} {7}},\ \bibinfo {pages} {527}
  (\bibinfo {year} {2011})}\BibitemShut {NoStop}%
\bibitem [{\citenamefont {Gieseler}\ \emph {et~al.}(2012)\citenamefont
  {Gieseler}, \citenamefont {Deutsch}, \citenamefont {Quidant},\ and\
  \citenamefont {Novotny}}]{Gieseler2012}%
  \BibitemOpen
  \bibfield  {author} {\bibinfo {author} {\bibfnamefont {J.}~\bibnamefont
  {Gieseler}}, \bibinfo {author} {\bibfnamefont {B.}~\bibnamefont {Deutsch}},
  \bibinfo {author} {\bibfnamefont {R.}~\bibnamefont {Quidant}}, \ and\
  \bibinfo {author} {\bibfnamefont {L.}~\bibnamefont {Novotny}},\ }\href
  {\doibase 10.1103/PhysRevLett.109.103603} {\bibfield  {journal} {\bibinfo
  {journal} {Phys. Rev. Lett.}\ }\textbf {\bibinfo {volume} {109}},\ \bibinfo
  {pages} {103603} (\bibinfo {year} {2012})}\BibitemShut {NoStop}%
\bibitem [{\citenamefont {Vovrosh}\ \emph {et~al.}()\citenamefont {Vovrosh},
  \citenamefont {Rashid}, \citenamefont {Hempston}, \citenamefont {Bateman},\
  and\ \citenamefont {Ulbricht}}]{Vovrosh}%
  \BibitemOpen
  \bibfield  {author} {\bibinfo {author} {\bibfnamefont {J.}~\bibnamefont
  {Vovrosh}}, \bibinfo {author} {\bibfnamefont {M.}~\bibnamefont {Rashid}},
  \bibinfo {author} {\bibfnamefont {D.}~\bibnamefont {Hempston}}, \bibinfo
  {author} {\bibfnamefont {J.}~\bibnamefont {Bateman}}, \ and\ \bibinfo
  {author} {\bibfnamefont {H.}~\bibnamefont {Ulbricht}},\ }\href@noop {}
  {}\bibinfo {note} {ArXiv:1603.02917}\BibitemShut {NoStop}%
\bibitem [{\citenamefont {Habraken}\ \emph {et~al.}(2012)\citenamefont
  {Habraken}, \citenamefont {Stannigel}, \citenamefont {Lukin}, \citenamefont
  {Zoller},\ and\ \citenamefont {Rabl}}]{Habraken2012}%
  \BibitemOpen
  \bibfield  {author} {\bibinfo {author} {\bibfnamefont {S.~J.~M.}\
  \bibnamefont {Habraken}}, \bibinfo {author} {\bibfnamefont {K.}~\bibnamefont
  {Stannigel}}, \bibinfo {author} {\bibfnamefont {M.~D.}\ \bibnamefont
  {Lukin}}, \bibinfo {author} {\bibfnamefont {P.}~\bibnamefont {Zoller}}, \
  and\ \bibinfo {author} {\bibfnamefont {P.}~\bibnamefont {Rabl}},\ }\href@noop
  {} {\bibfield  {journal} {\bibinfo  {journal} {New J. Phys.}\ }\textbf
  {\bibinfo {volume} {14}},\ \bibinfo {pages} {115004} (\bibinfo {year}
  {2012})}\BibitemShut {NoStop}%
\bibitem [{\citenamefont {Schmidt}\ \emph {et~al.}(2012)\citenamefont
  {Schmidt}, \citenamefont {Ludwig},\ and\ \citenamefont
  {Marquardt}}]{Schmidt2012}%
  \BibitemOpen
  \bibfield  {author} {\bibinfo {author} {\bibfnamefont {M.}~\bibnamefont
  {Schmidt}}, \bibinfo {author} {\bibfnamefont {M.}~\bibnamefont {Ludwig}}, \
  and\ \bibinfo {author} {\bibfnamefont {F.}~\bibnamefont {Marquardt}},\
  }\href@noop {} {\bibfield  {journal} {\bibinfo  {journal} {New J. Phys.}\
  }\textbf {\bibinfo {volume} {14}},\ \bibinfo {pages} {125005} (\bibinfo
  {year} {2012})}\BibitemShut {NoStop}%
\bibitem [{\citenamefont {Spreeuw}\ \emph {et~al.}(1990)\citenamefont
  {Spreeuw}, \citenamefont {van Druten}, \citenamefont {Beijersbergen},
  \citenamefont {Eliel},\ and\ \citenamefont {Woerdman}}]{Spreeuw1990}%
  \BibitemOpen
  \bibfield  {author} {\bibinfo {author} {\bibfnamefont {R.~J.~C.}\
  \bibnamefont {Spreeuw}}, \bibinfo {author} {\bibfnamefont {N.~J.}\
  \bibnamefont {van Druten}}, \bibinfo {author} {\bibfnamefont {M.~W.}\
  \bibnamefont {Beijersbergen}}, \bibinfo {author} {\bibfnamefont {E.~R.}\
  \bibnamefont {Eliel}}, \ and\ \bibinfo {author} {\bibfnamefont {J.~P.}\
  \bibnamefont {Woerdman}},\ }\href {\doibase 10.1103/PhysRevLett.65.2642}
  {\bibfield  {journal} {\bibinfo  {journal} {Phys. Rev. Lett.}\ }\textbf
  {\bibinfo {volume} {65}},\ \bibinfo {pages} {2642} (\bibinfo {year}
  {1990})}\BibitemShut {NoStop}%
\bibitem [{\citenamefont {Okamoto}\ \emph {et~al.}(2013)\citenamefont
  {Okamoto}, \citenamefont {Gourgout}, \citenamefont {Chang}, \citenamefont
  {Onomitsu}, \citenamefont {Mahboob}, \citenamefont {Chang},\ and\
  \citenamefont {Yamaguchi}}]{Okamoto2013}%
  \BibitemOpen
  \bibfield  {author} {\bibinfo {author} {\bibfnamefont {H.}~\bibnamefont
  {Okamoto}}, \bibinfo {author} {\bibfnamefont {A.}~\bibnamefont {Gourgout}},
  \bibinfo {author} {\bibfnamefont {C.-Y.}\ \bibnamefont {Chang}}, \bibinfo
  {author} {\bibfnamefont {K.}~\bibnamefont {Onomitsu}}, \bibinfo {author}
  {\bibfnamefont {I.}~\bibnamefont {Mahboob}}, \bibinfo {author} {\bibfnamefont
  {E.~Y.}\ \bibnamefont {Chang}}, \ and\ \bibinfo {author} {\bibfnamefont
  {H.}~\bibnamefont {Yamaguchi}},\ }\href@noop {} {\bibfield  {journal}
  {\bibinfo  {journal} {Nature Phys.}\ }\textbf {\bibinfo {volume} {9}},\
  \bibinfo {pages} {480} (\bibinfo {year} {2013})}\BibitemShut {NoStop}%
\bibitem [{\citenamefont {Faust}\ \emph {et~al.}(2013)\citenamefont {Faust},
  \citenamefont {Rieger}, \citenamefont {Seitner}, \citenamefont {Kotthaus},\
  and\ \citenamefont {Weig}}]{Faust2013}%
  \BibitemOpen
  \bibfield  {author} {\bibinfo {author} {\bibfnamefont {T.}~\bibnamefont
  {Faust}}, \bibinfo {author} {\bibfnamefont {J.}~\bibnamefont {Rieger}},
  \bibinfo {author} {\bibfnamefont {M.~J.}\ \bibnamefont {Seitner}}, \bibinfo
  {author} {\bibfnamefont {J.~P.}\ \bibnamefont {Kotthaus}}, \ and\ \bibinfo
  {author} {\bibfnamefont {E.~M.}\ \bibnamefont {Weig}},\ }\href@noop {}
  {\bibfield  {journal} {\bibinfo  {journal} {Nature Phys.}\ }\textbf {\bibinfo
  {volume} {9}},\ \bibinfo {pages} {485} (\bibinfo {year} {2013})}\BibitemShut
  {NoStop}%
\bibitem [{\citenamefont {Okamoto}\ \emph {et~al.}(2016)\citenamefont
  {Okamoto}, \citenamefont {Schilling}, \citenamefont {Sch\"{u}tz},
  \citenamefont {Sudhir}, \citenamefont {Wilson}, \citenamefont {Yamaguchi},\
  and\ \citenamefont {Kippenberg}}]{Okamoto2016}%
  \BibitemOpen
  \bibfield  {author} {\bibinfo {author} {\bibfnamefont {H.}~\bibnamefont
  {Okamoto}}, \bibinfo {author} {\bibfnamefont {R.}~\bibnamefont {Schilling}},
  \bibinfo {author} {\bibfnamefont {H.}~\bibnamefont {Sch\"{u}tz}}, \bibinfo
  {author} {\bibfnamefont {V.}~\bibnamefont {Sudhir}}, \bibinfo {author}
  {\bibfnamefont {D.~J.}\ \bibnamefont {Wilson}}, \bibinfo {author}
  {\bibfnamefont {H.}~\bibnamefont {Yamaguchi}}, \ and\ \bibinfo {author}
  {\bibfnamefont {T.~J.}\ \bibnamefont {Kippenberg}},\ }\href {\doibase
  10.1063/1.4945741} {\bibfield  {journal} {\bibinfo  {journal} {Appl. Phys.
  Lett.}\ }\textbf {\bibinfo {volume} {108}},\ \bibinfo {pages} {153105}
  (\bibinfo {year} {2016})}\BibitemShut {NoStop}%
\bibitem [{\citenamefont {Xu}\ \emph {et~al.}(2016)\citenamefont {Xu},
  \citenamefont {Mason}, \citenamefont {Jiang},\ and\ \citenamefont
  {Harris}}]{Xu2016}%
  \BibitemOpen
  \bibfield  {author} {\bibinfo {author} {\bibfnamefont {H.}~\bibnamefont
  {Xu}}, \bibinfo {author} {\bibfnamefont {D.}~\bibnamefont {Mason}}, \bibinfo
  {author} {\bibfnamefont {L.}~\bibnamefont {Jiang}}, \ and\ \bibinfo {author}
  {\bibfnamefont {J.~G.~E.}\ \bibnamefont {Harris}},\ }\href {\doibase
  doi:10.1038/nature18604} {\bibfield  {journal} {\bibinfo  {journal} {Nature}\
  } (\bibinfo {year} {2016}),\ doi:10.1038/nature18604}\BibitemShut {NoStop}%
\bibitem [{\citenamefont {Kiesel}\ \emph {et~al.}(2013)\citenamefont {Kiesel},
  \citenamefont {Blaser}, \citenamefont {Deli\'{c}}, \citenamefont {Grass},
  \citenamefont {Kaltenbaek},\ and\ \citenamefont {Aspelmeyer}}]{Kiesel2013}%
  \BibitemOpen
  \bibfield  {author} {\bibinfo {author} {\bibfnamefont {N.}~\bibnamefont
  {Kiesel}}, \bibinfo {author} {\bibfnamefont {F.}~\bibnamefont {Blaser}},
  \bibinfo {author} {\bibfnamefont {U.}~\bibnamefont {Deli\'{c}}}, \bibinfo
  {author} {\bibfnamefont {D.}~\bibnamefont {Grass}}, \bibinfo {author}
  {\bibfnamefont {R.}~\bibnamefont {Kaltenbaek}}, \ and\ \bibinfo {author}
  {\bibfnamefont {M.}~\bibnamefont {Aspelmeyer}},\ }\href {\doibase
  10.1073/pnas.1309167110} {\bibfield  {journal} {\bibinfo  {journal} {Proc.
  Natl. Acad. Sci. USA}\ }\textbf {\bibinfo {volume} {110}},\ \bibinfo {pages}
  {14180} (\bibinfo {year} {2013})}\BibitemShut {NoStop}%
\bibitem [{\citenamefont {Millen}\ \emph {et~al.}(2015)\citenamefont {Millen},
  \citenamefont {Fonseca}, \citenamefont {Mavrogordatos}, \citenamefont
  {Monteiro},\ and\ \citenamefont {Barker}}]{Millen2015}%
  \BibitemOpen
  \bibfield  {author} {\bibinfo {author} {\bibfnamefont {J.}~\bibnamefont
  {Millen}}, \bibinfo {author} {\bibfnamefont {P.~Z.~G.}\ \bibnamefont
  {Fonseca}}, \bibinfo {author} {\bibfnamefont {T.}~\bibnamefont
  {Mavrogordatos}}, \bibinfo {author} {\bibfnamefont {T.~S.}\ \bibnamefont
  {Monteiro}}, \ and\ \bibinfo {author} {\bibfnamefont {P.~F.}\ \bibnamefont
  {Barker}},\ }\href {\doibase 10.1103/PhysRevLett.114.123602} {\bibfield
  {journal} {\bibinfo  {journal} {Phys. Rev. Lett.}\ }\textbf {\bibinfo
  {volume} {114}},\ \bibinfo {pages} {123602} (\bibinfo {year}
  {2015})}\BibitemShut {NoStop}%
\bibitem [{\citenamefont {Larson}\ \emph {et~al.}(1986)\citenamefont {Larson},
  \citenamefont {Bergquist}, \citenamefont {Bollinger}, \citenamefont {Itano},\
  and\ \citenamefont {Wineland}}]{Larson1986}%
  \BibitemOpen
  \bibfield  {author} {\bibinfo {author} {\bibfnamefont {D.~J.}\ \bibnamefont
  {Larson}}, \bibinfo {author} {\bibfnamefont {J.~C.}\ \bibnamefont
  {Bergquist}}, \bibinfo {author} {\bibfnamefont {J.~J.}\ \bibnamefont
  {Bollinger}}, \bibinfo {author} {\bibfnamefont {W.~M.}\ \bibnamefont
  {Itano}}, \ and\ \bibinfo {author} {\bibfnamefont {D.~J.}\ \bibnamefont
  {Wineland}},\ }\href {\doibase 10.1103/PhysRevLett.57.70} {\bibfield
  {journal} {\bibinfo  {journal} {Phys. Rev. Lett.}\ }\textbf {\bibinfo
  {volume} {57}},\ \bibinfo {pages} {70} (\bibinfo {year} {1986})}\BibitemShut
  {NoStop}%
\bibitem [{\citenamefont {J\"{o}ckel}\ \emph {et~al.}(2015)\citenamefont
  {J\"{o}ckel}, \citenamefont {Faber}, \citenamefont {Kampschulte},
  \citenamefont {Korppi}, \citenamefont {Rakher},\ and\ \citenamefont
  {Treutlein}}]{Joeckel2015}%
  \BibitemOpen
  \bibfield  {author} {\bibinfo {author} {\bibfnamefont {A.}~\bibnamefont
  {J\"{o}ckel}}, \bibinfo {author} {\bibfnamefont {A.}~\bibnamefont {Faber}},
  \bibinfo {author} {\bibfnamefont {T.}~\bibnamefont {Kampschulte}}, \bibinfo
  {author} {\bibfnamefont {M.}~\bibnamefont {Korppi}}, \bibinfo {author}
  {\bibfnamefont {M.~T.}\ \bibnamefont {Rakher}}, \ and\ \bibinfo {author}
  {\bibfnamefont {P.}~\bibnamefont {Treutlein}},\ }\href@noop {} {\bibfield
  {journal} {\bibinfo  {journal} {Nat. Nanotech.}\ }\textbf {\bibinfo {volume}
  {10}},\ \bibinfo {pages} {55} (\bibinfo {year} {2015})}\BibitemShut {NoStop}%
\bibitem [{\citenamefont {Jain}\ \emph {et~al.}(2016)\citenamefont {Jain},
  \citenamefont {Gieseler}, \citenamefont {Moritz}, \citenamefont {Dellago},
  \citenamefont {Quidant},\ and\ \citenamefont {Novotny}}]{Jain2016}%
  \BibitemOpen
  \bibfield  {author} {\bibinfo {author} {\bibfnamefont {V.}~\bibnamefont
  {Jain}}, \bibinfo {author} {\bibfnamefont {J.}~\bibnamefont {Gieseler}},
  \bibinfo {author} {\bibfnamefont {C.}~\bibnamefont {Moritz}}, \bibinfo
  {author} {\bibfnamefont {C.}~\bibnamefont {Dellago}}, \bibinfo {author}
  {\bibfnamefont {R.}~\bibnamefont {Quidant}}, \ and\ \bibinfo {author}
  {\bibfnamefont {L.}~\bibnamefont {Novotny}},\ }\href {\doibase
  10.1103/PhysRevLett.116.243601} {\bibfield  {journal} {\bibinfo  {journal}
  {Phys. Rev. Lett.}\ }\textbf {\bibinfo {volume} {116}},\ \bibinfo {pages}
  {243601} (\bibinfo {year} {2016})}\BibitemShut {NoStop}%
\bibitem [{\citenamefont {Frimmer}\ and\ \citenamefont
  {Novotny}(2014)}]{Frimmer2014}%
  \BibitemOpen
  \bibfield  {author} {\bibinfo {author} {\bibfnamefont {M.}~\bibnamefont
  {Frimmer}}\ and\ \bibinfo {author} {\bibfnamefont {L.}~\bibnamefont
  {Novotny}},\ }\href@noop {} {\bibfield  {journal} {\bibinfo  {journal} {Am.
  J. Phys.}\ }\textbf {\bibinfo {volume} {82}},\ \bibinfo {pages} {947}
  (\bibinfo {year} {2014})}\BibitemShut {NoStop}%
\end{thebibliography}%

\end{document}